# Organizational adaptation to Complexity: A study of the South African Insurance Market as a Complex Adaptive System through Statistical Risk Analysis

Satyakama Paul, Bhekisipho Twala, Tshilidzi Marwala
Department of Mechanical Engineering Science, FEBE B2 Lab 210, Auckland Park Kingsway Campus,
University of Johannesburg, Johannesburg, 2006, South Africa

**Abstract** South Africa assumes a significant position in the insurance landscape of Africa. The present research based upon qualitative and quantitative analysis, shows that it shows the characteristics of a Complex Adaptive System. In addition, a statistical analysis of risk measures through Value at risk and Conditional tail expectation is carried out to show how an individual insurance company copes under external complexities. The authors believe that an explanation of the coping strategies, and the subsequent managerial implications would enrich our understanding of complexity in business.

## 1. Introduction

South Africa (SA) assumes an important place in the insurance landscape of the African continent, contributing almost three quarter of the total African insurance. Developing at a compound annual growth rate (CAGR) of 11.7% in between 2004-08, the value of the market measured in gross premium income was USD 3.7 billion in 2008 [6]. The life insurance segment is expected to grow at a CAGR of 8.3% in between 2009-14 [5], and the non-life segment at 2.8% in between 2008-13 [7].

The present research is aimed at understanding the automobile insurance market of SA with a focus upon certain market level and company level features. Although the performance of any segment of the insurance market depends upon a range of economic and non-economic factors, yet till data very little effort has been channelized at studying it from the Complexity point of view. Though qualitative and quantitative analysis of industry and company data, the authors argue that the SA automobile insurance market can be viewed as a Complex Adaptive System (CAS), and the organizational adaptations to external complexity can provide considerable insight into the behaviour of the overall market.

In addition, while considerable research has been directed at studying other industries as a CAS - example: Software development [16]; Military organizations [10]; Health care [11], [22], [1]; Manufacturing and logistics [18], Education [19] etc, limited research have been carried to date in understanding financial and specifically insurance companies as CAS. So in this regard, a novel endeavour is carried in studying an automobile insurance company as CAS. The authors believe that such an effort would further enrich our understanding of complexity in business.

## 2. Complex Adaptive System Theory and the Insurance Market

Jervis defines a system is "a set of units or elements … interconnected so that changes in some elements or their relations produce changes in other parts of the system" [12], cited in [15]. Sanders and McCabe tells us that "Complex adaptive systems, and models thereof, are characterized by distributed organizations or networks, whose parts all influence each other, either directly or through feedback loops, which continually evolve and adapt to accomplish overarching goals"[23]. In an effort to explain why the SA insurance market can be viewed as a CAS, first, we review similar works (in other sectors) of authors as Pohl[20], Chan[3], Kelly[13], Zhao[25], Meso and Jain[16], Nilsson and Darley[18], Schneider and Somers[24], Palmberg[19] etc. Second, we compare the summarised characteristics (following italic words) of CAS with various macro (i.e. market) and micro (i.e. company) level features of the insurance market to logically conclude that the market is a CAS.

The agents of the insurance market are the insurance firms, individual and corporate policy holders, Government regulatory bodies, and Information Technology (IT) manufacturers[1]. These agents are entwined in a *network structure* and interact with other agents in its immediate vicinity by *simple localized rules*. For example an individual buying an automobile insurance, for most times, in addition to insurance prices, ends up buying a policy based upon his perception of the company, opinion of neighbours, reviews in blogs etc. On the other hand, the intense rivalry and price wars between the insurance firms restrict the free flow of information [6]. Thus in bid to retain high-margin customers, an individual firm may provide an unnaturally large loyalty discounts that is based upon its imperfect information of the market. It might be worth mentioning here that arguing upon the 2008 data [7], the SA non-life insurance market is not a perfectly competitive market. While the largest shareholder Santam Ltd. held 23% share, Mutual & Federal - 15.8%, Holland - 9.6%, and others held the rest 51.6%.

---
[1] The IT manufacturers provide the insurance companies with specialized and often custom made software for their business operations.

The insurance market in SA is speedily changing. With rapid changes in the behaviour of policy holders - brand image of the company assumes more importance than ever before [14], growing consumerism and better financial decisions of certain customer segments, growing conflicts in the interests of the policy holders and shareholders [2] etc., the Financial Services Board (FSB)[2] of SA have increasingly felt the need to move from Basel II to Solvency Assessment and Management (SAM) regime. SAM promises better tools for risk monitoring and management for the insurers by allowing them to develop "…full and partial internal capital models and increased use of risk mitigation and risk transfer tools"[8], take more proactive responses to mitigate the uncertainties of the market, and provide better alignment between the interests of all stakeholders [14]. This might serve as a good example of *change* and *adaptability* in CAS – in anticipating the future - the agents (FSB and the insurers) of a CAS are always interacting with one another in its immediate vicinity. Richardson [21] terms this autonomy of interaction of the agents as their 'local memory'. As the agents with local memory constantly learn from their newer experiences (changing consumer behaviour, technology, etc), re-organize themselves in accordance with the changing environment, the CAS changes constantly (embracing SAM), and gets adapted to new, unexpected conditions [26].

In order to show *non-linearity* of CAS, we take the example of an individual SA automobile insurance company. The company provided datasheets contained information of approximately 22 thousand policy holders - information about their vehicle (type and age); and person level characteristics (age, gender, prior driving experience, etc); type of loss from accident – (1) losses from injury to a party other than the insured, (2) losses for damages to the insured, including injury, property damage, fire and theft, and (3) losses for property damage to a party other than the insured. It is to be noted that occasionally there may be more than one type of loss incurred with each accident.

We use the negative binomial probability distribution to calculate the predicted mean for each level of No Claims Discount (NCD) against the random variables that are the possible causes of an accident[3] - Individual Loss, Sum of Losses from a Type, and Sum of Losses from a Specific Event. It is to be noted here that NCD is the discount (on insurance premium) that a driver receives from the insurer for a previous claim free driving history. Also for our analysis, the level of NCD is calculated by assuming that at NCD equal to one hundred, a driver will receive a discount equal to his total premium from the insurer; and at a NCD equal to zero, he can will receive zero discount.

Individual Loss, Sum of Losses from a Type, and Sum of Losses from a Specific Event are respectively referred to as Indl. Loss, ∑Losses-Type, and ∑Losses-Sp. Event in table 1 and figure 1. The negative ordinate axis of Figure 1 shows that for proportionate change (of 10) in the level of NCD, and the positive abscissa shows the predicted mean of the three random variables. Figure 1 shows that the predicted means of the three random variables may either show no change, less than or more than proportionate increase or decrease, and the amount of change is unpredictable. Thus we can infer that there exists no "causal relationship" [4] between the interactions of the agents and the outcome – hence the company exhibits the *non-linearity* characteristic of CAS.

Table 1. Predictive Mean by Level of NCD

| Type of Random Variable | Level of NCD | | | | | |
|---|---|---|---|---|---|---|
| | 0 | 10 | 20 | 30 | 40 | 50 |
| Indl. Loss | 230.7 | 205.7 | 167.8 | 163.3 | 147.8 | 121.7 |
| ∑Losses-Type | 336 | 291.5 | 239.6 | 223.4 | 208.7 | 169.8 |
| ∑Losses-Sp. Event | 385.6 | 358.9 | 313.4 | 304.6 | 289.4 | 235.6 |

---

[2] The FSB is an independent institution, set up by statute to supervise the SA Non-banking Financial Services Industry in the interest of the public.
[3] The rationale behind the use of the negative binomial distribution lies in the fact that accident and NCD are mutually exclusive to one another.

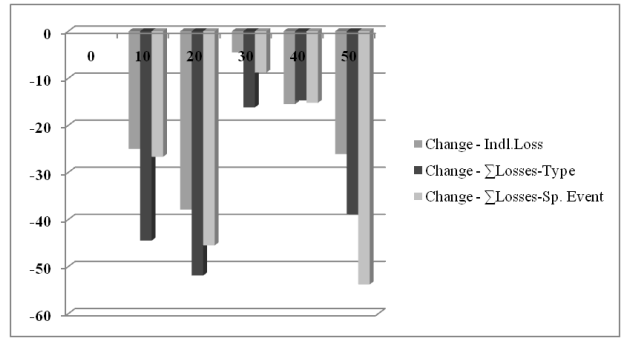

Fig 1. Change of Predictive Mean for equal changes in Level of NCD

Lastly, we show the automobile insurance company manifests the '*edge of chaos*" – a point of dynamic equilibrium. Quota share reinsurance is a form of proportional reinsurance which specifies that a fixed percentage of each policy written will be transferred to the reinsurer. The effect of different quota shares on the retained claims for the ceding company is examined and presented in figure 2.

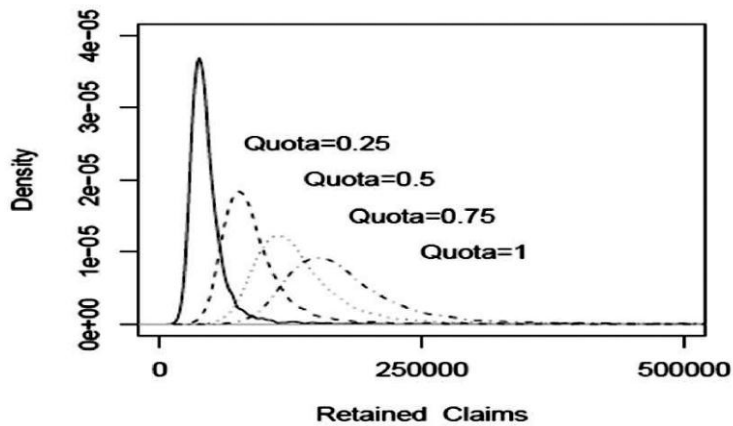

Fig. 2. Distribution of retained claims for the insurer under quota share reinsurance

The distributions of retained claims are derived assuming the insurer retains 25%, 50%, 75% and 100% of its business. In figure 2, a quota of 0.25 means the insurer retains 25% of losses and cedes 75% to the reinsurer. The curve corresponding to a quota of 1 represents the losses of insurer without a reinsurance agreement. As we can see, the quota share reinsurance does not change the shape of the retained losses, only the location and scale. For example, if the insurer ceded 75% of losses, the retained losses will shift left and the variance of retained losses will be 1/16 times the variance of original losses. Thus a decrease in the variance means that - with a continuous change in the environment (i.e. continuous interaction between the insurer and reinsurer regarding sharing of claims); a point of dynamic equilibrium is reached. The arguments in the above paragraphs prove that the SA automobile industry shows all the major characteristics of a CAS.

## 3. Organizational Adaptation to External Complexity

In this section we try to analyse the effects of external complexity on the above-mentioned organization. Following the works of Milling [17] and Großler et al.[9], we understand complexity to consist of detailed complexity and dynamic complexity. Detailed complexity consists of the number of agents in the system, the connections between them, and the functional relationship that binds the agents. Dynamic complexity explains the change of the structure of the system over time.

Großler et al. [9] work in the manufacturing industries tells us that in high complexity environments, organizations – in order to stay competitive, improve their performance in three strategic fronts (a) cost, (b) quality, and (c) flexibility. In this present example, we show similar results can be observed in the insurance sector (in our case: the automobile insurance company). Thus our research hypothesis is:

*H1: An automobile insurance company under high external complexity would adapt internally to incorporate the strategy of 'increase in flexibility' and 'decrease in cost' with regard to its selection of coverage's arising from the claims.*

Other than the various aspects of the market (already discussed above) that make it a CAS, we now concentrate on the complexity of the insurance claim. We view an insurance claim as a product. As shown earlier, the loss incurred by an insurer from the coverage of a claim may be of three types: (i) actual losses from third party injury, (ii) actual losses from own damage (of the driver), and (iii) actual losses from third party property. Following Großler et al [9] rationale that the complexity of a product depends upon the design of the product, we can conveniently assume that insurance claim is a complex product as losses from it depends upon the above mentioned three factors. The distributions for actual losses for the three types of losses are shown in Figure 3. The measures on the three vertical axis's also show the significant differences among densities.

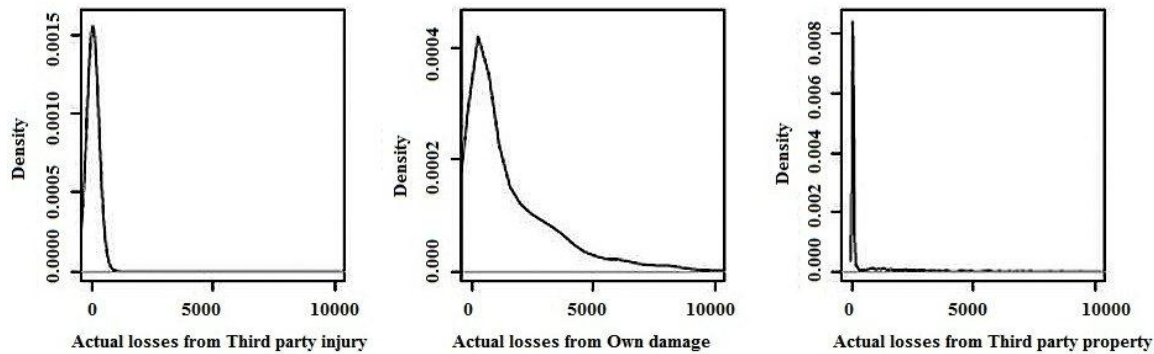

Fig. 3. Distribution of Losses by Claim Type

This type of analysis can provide useful information about the risk profile that arises from the losses due to different coverage's. An analysis of the above figures show that, within a loss of approximately five thousand in claim, the insurer has the highest probability of losing money from third party injury, followed by third party property, and lastly from own damage. Thus an insurance analyst may consider which type of coverage should be incorporated and which type should be eliminated, so that the product can be tuned to meet the company's risk management requirements, regulatory policies, and other strategic goals. The need for incorporation or elimination of any type of coverage is an example of need for increased flexibility arising out of product customization. Similar results are observed by Großler et al. [9] in their research on the manufacturing industry.

Another way to examine the role of dependence is to decompose the comprehensive coverage into more "primitive" coverage's for the three types of claims. As in derivative securities, we call this "unbundling" of coverage's. We are able to calculate risk measures for each unbundled coverage, as if separate financial institutions owned each coverage, and compare them to risk measures for the bundled coverage that the insurance company is responsible for. It is to be noted here that risk can be viewed as a cost to the insurance company. The results are shown in Table 2.

Table 2. Value at risk (VaR) and Conditional tail expectation (CTE) by Percentile for Unbundled and Bundled Coverage's

|  | VaR | | | CTE | | |
|---|---|---|---|---|---|---|
| Unbundled Coverage's | 90% | 95% | 99% | 90% | 95% | 99% |
| Third party injury | 61.4 | 209.8 | 1063.9 | 492.3 | 864.2 | 2557.9 |
| Own damage | 4.9 | 5.9 | 8.6 | 6.5 | 7.6 | 10.4 |
| Third party property | 88.2 | 109.5 | 164.3 | 123.5 | 148.8 | 224.3 |
| Sum of Unbundled Coverage's | 299.3 | 479.2 | 1415.1 | 781.4 | 1190.1 | 2086.7 |
| Bundled (Comprehensive) Coverage | 158.6 | 224.6 | 663.2 | 368.3 | 552.3 | 1437.7 |

The risk measures for bundled coverage's are smaller than the sum of unbundled coverage's, for both risk measures and all percentiles. One of the important purposes of risk measures is to determine economic capital which is the amount of capital that banks and insurance companies set aside as a buffer against potential losses from extreme risk event. The implication of the above table is that by bundling different types of coverage into one comprehensive policy, the insurers can reduce the economic capital for risk management or regulatory purpose. Another perspective is that this example simply demonstrates the effectiveness of economies of scales; three small financially independent institutions (one for each coverage) require in total more capital than a single combined institution (one for the bundled coverage).

The reference price experiment we conducted proves our hypothesis that people do not like unbundled pricing because their reference price for these was R0.00[4] but can be nudged towards unbundled pricing if their reference price can be improved. But what if the insurer who offered unbundled option wants to shift the customers to the new bundled option? How should the choice design be?

In this reverse unbundling case there are two possible scenarios

1. The bundled product is the sum of its parts – no other value-adds. In this case the price of the bundle should less than or equal to the unbundled option as previously discussed.

2. The bundled product has a new component that was not offered before and is not made available in the unbundled option. In this case the price of the bundle can be more than the sum of the prices of unbundled components, if and only if the value proposition of the new components are clear to the customers and they have a non-zero reference price.

In the second case, if the value communication is not clear then customers will be highly unlikely to pay the premium for the bundled option. More importantly, if the reference price for the new value-added component is R0.00, then customers will be more than likely to prefer the unbundled option.

So in both unbundling and bundling cases, reference price is key. If the insurer wants to capture part of the value-added they should focus on setting a higher reference price in the minds of their customers.

## 4 Conclusion and Managerial Implications

Initially the paper set out to show that the insurance market of SA is a CAS. Through qualitative and quantitative analysis, the second section of the paper successfully shows how the automobile insurance market and an insurance company operating in it show the important characteristics of a CAS. This would have an important implication for the insurance industry professionals. Since there exists no linear relationship between the interactions of the agents and the outcome produced by the system, so the principles of modern management - "*stability as an objective, analysis by reduction to parts, and cause and effect mechanisms between the parts*"[19] might not well characterize the SA insurance market. In addition the conventional policy making tools based upon the interaction of demand and supply might not well bring in the desired results. The discussions in section 2 also hint at an important conclusion: today's insurance market customer other than the price of a product is strongly influenced by a lot of other environmental factors. Thus a single optimization objective (based upon profit, sales etc) may not be a sound policy of an insurance company.

The empirical analysis in the third section shows that in case of external complexity, an automobile company would adapt internally by increasing the flexibility in its product and make it more customized to suit the organization's risk requirements, and decrease cost. The results obtained are similar to (Großler et al.[9] for manufacturing industries. It also shows the importance that a reinsurer plays in reducing the risk cost (and hence increase the profit) of an insurance company. Thus a strong strategic relationship between an insurer and reinsurer is foreseen.

---

[4] R stands for Rand, South African currency.